\documentclass{elsarticle}
\usepackage{graphicx}
\usepackage{amssymb}
\usepackage{amsmath}
\usepackage{amsfonts}

\begin{document}

\begin{frontmatter}
\title{Recent results of the ANTARES Neutrino Telescope}
\author{Juan Jos\'e Hern\'andez-Rey\\(for the ANTARES Collaboration)}
\address{IFIC - Instituto de F\'{\i}sica Corpuscular\\  
Universitat de Val\`encia--CSIC,\\
E-46100 Valencia, Spain}
\begin{abstract}
 The latest results from the ANTARES Neutrino Telescope are reported.
 Limits on a high energy neutrino diffuse flux have been set using for
 the first time both muon--track and showering events.  The results
 for point sources obtained by ANTARES are also shown. These are the
 most stringent limits for the southern sky for neutrino energies
 below 100~TeV.  Constraints on the nature of the cluster of neutrino
 events near the Galactic Centre observed by IceCube are also
 reported.  In particular, ANTARES data excludes a single point--like
 neutrino source as the origin of this cluster.  Looking for neutrinos
 coming from the Sun or the centre of the Galaxy, very competitive
 limits are set by the ANTARES data to the flux of neutrinos produced
 by self-annihilation of weakly interacting massive particles.
\end{abstract}
\end{frontmatter}

\section{Introduction}
Several astrophysical objects both Galactic and extra-galactic have
been proposed as sites of acceleration of protons and nuclei, but no
conclusive experimental evidence has been obtained yet and in any case
in-depth experimental studies of the cosmic hadronic accelerators are
lacking.  The decays of mesons produced by the interactions of protons
and nuclei with matter or radiation would yield neutrinos, thus
indicating the presence of this type of acceleration. The detection of
high energy cosmic neutrinos can therefore shed light on the origin of
cosmic rays. Furthermore, neutrinos are at the end of a variety of
decay chains of standard (and beyond the standard) model particles,
being therefore an exceedingly useful ``debris'' to look for different
processes, such as for instance the self-annihilation of the
hypothetical weakly interacting massive particles that could form the
dark matter in the Universe.

Let us briefly summarize the advantages of neutrinos as cosmic
messengers. They are neutral particles, therefore they are not
deflected by magnetic fields and point back to their sources. They are
weakly interacting and thus can escape from very dense astrophysical
objects and travel long distances without being absorbed by matter or
background radiation.  Moreover, in cosmic sites where hadrons are
accelerated, neutrinos are generated in the decay of charged pions
produced in the interaction of those hadrons with the surrounding
matter or radiation, being therefore a smoking gun of hadronic
acceleration mechanisms.

Several neutrino telescopes are at present operating worldwide and
larger telescopes or extensions of the already existing are planned.
The breakthrough in this field took place in 2013 with the
announcement by the IceCube collaboration of the first evidence of a
cosmic signal of high energy neutrinos~\cite{HESE1} with the
subsequent confirmation of the signal with more data~\cite{HESE2}.

Here we report the recent results of the ANTARES neutrino telescope.
Even though of a much smaller size than IceCube, ANTARES is capable of
providing useful information both in the search of neutrino
astrophysical sources and that of indirect dark matter, as we show in
this contribution.

\section{The ANTARES telescope}
The ANTARES Collaboration completed the construction of its namesake
neutrino telescope in the Mediterranean Sea in May 2008, although a
partial version of the device was operating since 2007.  The
telescope, located 40 km off the southern coast of France
(42$^{\circ}$48'N, 6$^{\circ}$10'E) at a depth between 2475 m (seabed)
and 2025 m (top of the lines), consists in a three-dimensional array
of photomultipliers housed in glass spheres, called optical modules,
distributed along twelve lines anchored to the sea bottom and kept
taut by a buoy at the top. Each line is composed of 25 storeys of
triplets of optical modules (OMs), each housing one 10-inch photomultiplier.
The lines are subjected to the sea currents and can change shape and
orientation. A positioning system based on hydrophones, compasses and
tiltmeters is used to monitor the detector geometry with an accuracy
of about $10$~cm. More details of the ANTARES telescope can be found
in ref.~\cite{Antares}.

The goal of the experiment is to search for neutrinos with energies
greater than $\sim$50~GeV mainly by detecting muons.  A muon neutrino
that has crossed the Earth can undergo a charged current interaction
before arriving to the detector and produce a muon that can travel
hundreds of metres and cross the telescope.  Muons induce the emission
of Cherenkov light in sea water and the arrival time and intensity of
this light on the OMs are digitized into hits and transmitted to
shore.  Events containing muons are selected from the continuous
deep--sea optical backgrounds due to natural radioactivity and
bioluminescence. The arrival time of the Cherenkov photons can be
determined at the nanosecond level~\cite{TimeCalib}, allowing the
measurement of the direction of upgoing tracks with resolutions better
than 0.5$^\circ$ for neutrino energies above 1~TeV.  Due to the large
background from downgoing atmospheric muons, the telescope is
optimised for the detection of upgoing muons that can only originate
from neutrinos. Recently, also neutrino-induced shower events are
being reconstructed increasing the reach of the detector in terms of
detectable neutrino types.

\section{Diffuse fluxes}

A search for a neutrino diffuse flux using upgoing muon neutrino
events has been performed using the data recorded from 2007 to 2011,
corresponding to a total livetime of 885 days (throughout this
contribution, the words \emph{neutrino} and \emph{muon} are meant to
include also their antiparticle). The analysis first imposes loose cuts
on the muon track reconstruction quality parameter and the angular
error estimate obtained by the reconstruction fit to reduce the
background. On this reduced sample, optimised cuts are then applied to
the quality parameter and the number of hits in the event, used as a
proxy of the neutrino energy, whose spectrum is expected to be harder
for the signal than for the background. The optimisation of the cuts
was carried out following a blinding procedure, i.e. on simulated
data and accessing only 10\% of the total data sample which was
subsequently discarded for the final analysis. After unblinding,
i.e. when applying the selection to the data, 8 events pass the cuts
while the expected background is 8.4 events.

\begin{figure} [ht] 
\centering
  \includegraphics[height=.4\textheight]{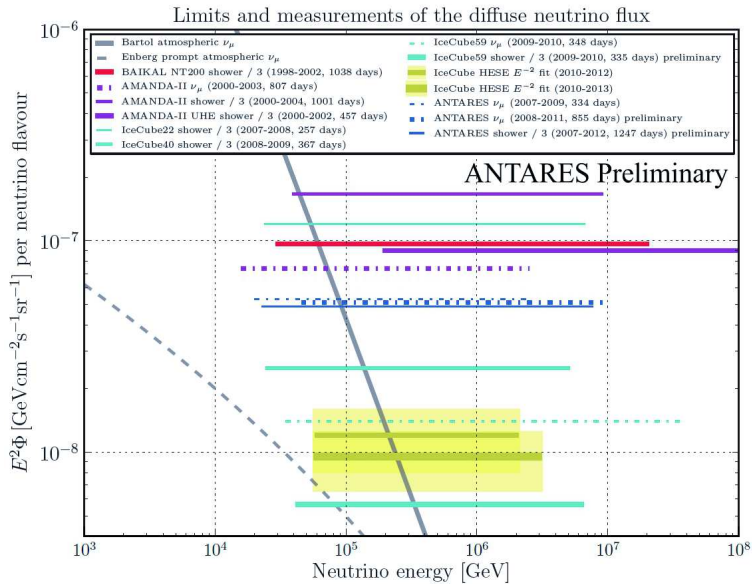}
  \caption{90\% C.L. upper limits to an E$^{-2}$ diffuse
    neutrino flux for different experiments and analyses. The
    ANTARES upper limits (blue lines) set by the muon neutrino and
    showering event analyses are 5.1 and 
  4.9~$\times$~10$^{-8}$~GeV~cm$^{-2}$~s$^{-1}$~sr$^{-1}$, 
  respectively.  See text for
    explanations.}
\label{diffuselimits}
\end{figure}

This corresponds to a 90\% confidence level upper limit (\`a la
Feldman-Cousins with systematic errors included):

\begin{equation}
 E^{-2} \cdot \, \Phi_{90\%} = 5.1 \times 10^{-8} \, \, 
\textrm{GeV cm}^{-2} \textrm{s}^{-1} \textrm{sr}^{-1}
\end{equation}
\noindent in the energy range from 45~TeV to 10~PeV. This limit is
shown in Fig.~\ref{diffuselimits}

\begin{figure}[ht]
\centering
  \includegraphics[height=.4\textheight]{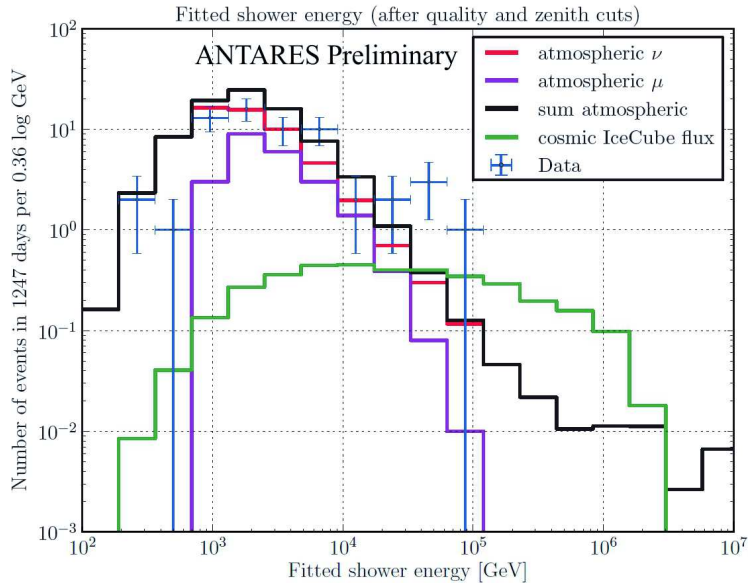}
  \caption{Distribution of the reconstructed energy of the selected
    shower events before the final energy cut. The data are the points with
    errors. The expected background contributions are also shown:
    atmospheric neutrinos (red line), atmospheric muons (purple line),
    sum of both (black line). Finally, how a flux of the magnitude of that
    observed by IceCube's high-energy starting events would look like 
    is also represented. See text for explanations. }
\label{showerenergy}
\end{figure}

\begin{figure}[ht]
\centering
  \includegraphics[height=.4\textheight]{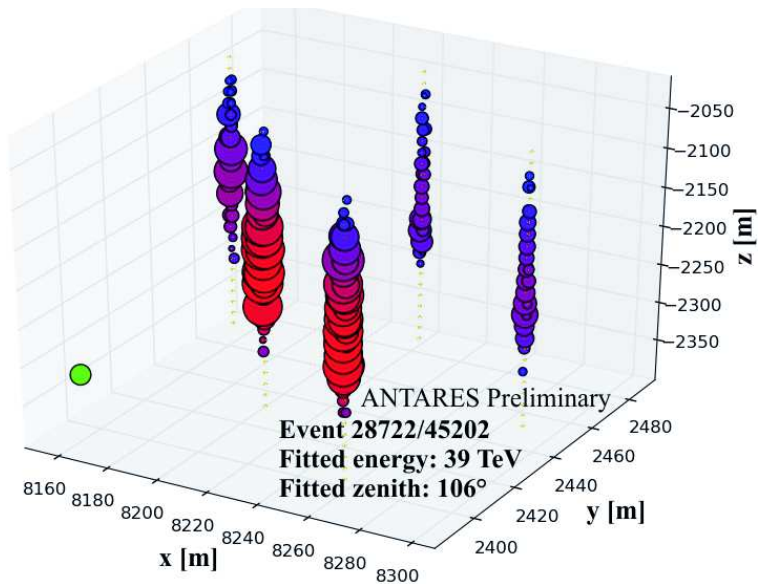}
  \caption{Example of shower event that passes the selection criteria.}
\label{showerevent}
\end{figure}

An alternative analysis was performed that included various event
parameters and an optimal set of event selection criteria were found
by scanning this parameter space. One of the parameters was the energy
of the event, which was estimated in this case by an artificial neural
network.  The livetime in this analysis was slightly higher, 900 days,
but the final sensitivity was very similar to the previous analysis:
4.2 $\times \, \Phi_0$ versus 4.7 $\times \, \Phi_0$ for the previous analysis
(where $\Phi_0 =$ 10$^{-8}$ GeV cm$^{-2}$ s$^{-1}$ sr$^{-1}$).  After
unblinding, 12 events pass the cuts on an expected background of 8.4
events. This small excess interpreted as a background fluctuation
gives a 90\% C.L. limit slightly higher but fully compatible with the
previous result, 7.7 $\times \, \Phi_0$ versus 5.1 x $\Phi_0$, in a
similar energy range (from 65~TeV to 10~PeV),

The search for a diffuse neutrino flux has also been performed using
shower events. The shower reconstruction algorithm first selects good
signal hits from all the recorded PMT signals in the event. From these
selected hits, the interaction vertex location and time are
reconstructed using a maximum likelihood method. From this vertex and
using the shower signal hits the energy and the neutrino direction are
estimated.

 After the muon suppression cut (see later), the median vertex and
 direction errors are 4 metres and 6$^{\circ}$, respectively, for 10 TeV
 showers, and the logarithm of the energy (log$(E_{fit}/E_{MC})$) is
 reconstructed with an  error of -0.16 for showers of 10 TeV. Of
 all the simulated showers, 40\% are reconstructed and pass the muon
 cut at 10~TeV and 90\% at 10~PeV. At this level, the muon rejection
 power is of the order of 10$^5$.

 Once the shower is reconstructed a series of selection criteria are
 imposed. First, a muon filter is used to reject those events
 that may be compatible with a muon track. The events are then
 required to be reconstructed in more than two lines. Although
 unfrequent, some optical modules produce from time to time sparks
 that could imitate a shower. A special filter based on a minimum
 distance of the reconstructed shower vertex to any OM is used to
 reject these events. Finally, the event is required to be upgoing
 (fitted zenith angle greater than 94$^{\circ}$) and the reconstructed
 energy of the shower should be higher than 10~TeV.

The expected background after this selection criteria is 4.9 events,
coming mainly from atmospheric neutrinos (3.1 events) and muons (1.9
events). The systematic error on the total background is $\pm$2.9
events.  In Fig.~\ref{showerenergy} the distribution in energy of all
the events before the final cut on reconstructed energy is shown.

After unblinding, 8 shower events pass all the cuts.  In
Fig.~\ref{showerevent}, one of the events is shown. The probability to
obtain 8 or more background events when 4.9 are expected is 12.5\%
(1.5$\sigma$). Therefore, this excess is interpreted as a background
fluctuation and the following 90\% confidence level upper limit on a
cosmic signal is extracted using Feldman--Cousins and taking into
account the systematic uncertainties:

\begin{equation}
 E^{-2} \cdot \, \Phi_{90\%} = 4.91 \times 10^{-8} \, \, 
\textrm{GeV cm}^{-2} \textrm{s}^{-1} \textrm{sr}^{-1}
\end{equation}
\noindent in the energy range from 23~TeV to 7.8 ~PeV. This limit is
shown in Fig.~\ref{diffuselimits}.

Given the cosmic signal observed in IceCube~\cite{HESE2}, namely:
$1.0\pm 0.3 \times \Phi_0$ (where again, $\Phi_0$ = 10$^{-8}$ GeV
cm$^{-2}$ s$^{-1}$sr$^{-1}$) with a hard cut-off at 1.6~PeV, one may
wonder to which flux the ANTARES excess of shower events would
correspond if interpreted as a cosmic signal. The answer is that with
an unbroken E$^{-2}$ spectrum the excess corresponds to a flux
intensity of $1.3 ^{+1.8}_{-1.3}$ $\times \, \Phi_0$, and to
1.7$^{+2.3}_{-1.8}$ $\times \, \Phi_0$, assuming a cut-off at 2 ~PeV.
Therefore, we cannot also exclude either that this excess is not a
background fluctuation, but lack of statistics prevent us from making
any claim.

\section{Search for point sources}
Although already published~\cite{antpoint}, we want to report briefly
on some of the results of the search for neutrino point sources
recently released by ANTARES, because of their impact on the recent
signal observed by the IceCube collaboration~\cite{HESE1, HESE2}. The
data used for this analysis were recorded between 2007 and 2012 and
correspond to a total livetime of 1338 days. Upgoing muon neutrino
events leaving a well-reconstructed muon track in the detector were
searched for.  The parameters used to select the events were the
reconstruction quality of the corresponding track, its angular
uncertainty as estimated by the fit and its zenith angle. The exact
values of these parameters were chosen so that the neutrino flux
required to make a 5$\sigma$ discovery with 50\% probability was
minimised. As usual, this minimisation was performed following a blind
procedure, i.e. using pseudo-experiments before performing the
analysis on the data.  After unblinding, the selection gave a total of
5516 events, which included an estimated background of 10\% of
misreconstructed atmospheric muons. Signal events are expected to
accumulate in clusters over the diffuse background of atmospheric
neutrinos.  The search for clusters is performed with a maximum
likelihood method fed with information about the angular error
estimate of the events and their energy via the number of observed
hits in the event. The minimisation provides a number of signal events
and a test statistic from which we can extract the probability of the
observation to be produced by the expected background ($p$-value).

\begin{figure}[ht]
\centering
  \includegraphics[height=0.5\textheight]{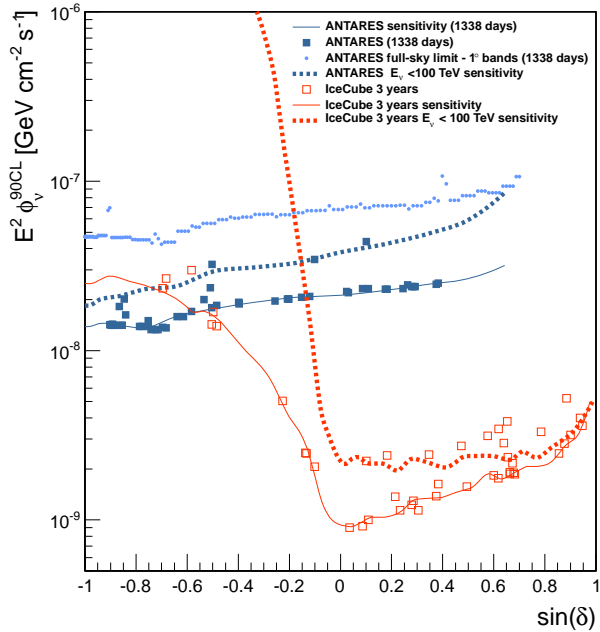}
  \caption{ 90\% C.L. flux upper limits and sensitivities on the muon neutrino
     flux for six years of ANTARES data. IceCube results are also
     shown for comparison. The light-blue markers show the upper limit
     for any point source located in the ANTARES visible sky in
     declination bands of 1$^{\circ}$. The solid blue (red) line indicates the
     ANTARES (IceCube) sensitivity for a point-source with an E$^{-2}$
     spectrum as a function of the declination. The blue (red)
     squares represent the upper limits for the ANTARES (IceCube)
     candidate sources. Finally, the dashed dark blue (red) line
     indicates the ANTARES (IceCube) sensitivity for a point-source
     and for neutrino energies lower than 100 TeV, which shows that
     the IceCube sensitivity for sources in the southern hemisphere
     is mostly due to events of higher energy. The IceCube results
     were derived from ref.~\cite{ICpoint}.}
\label{psources}
\end{figure}

\begin{figure}[ht]
\centering
  \includegraphics[height=0.45\textheight]{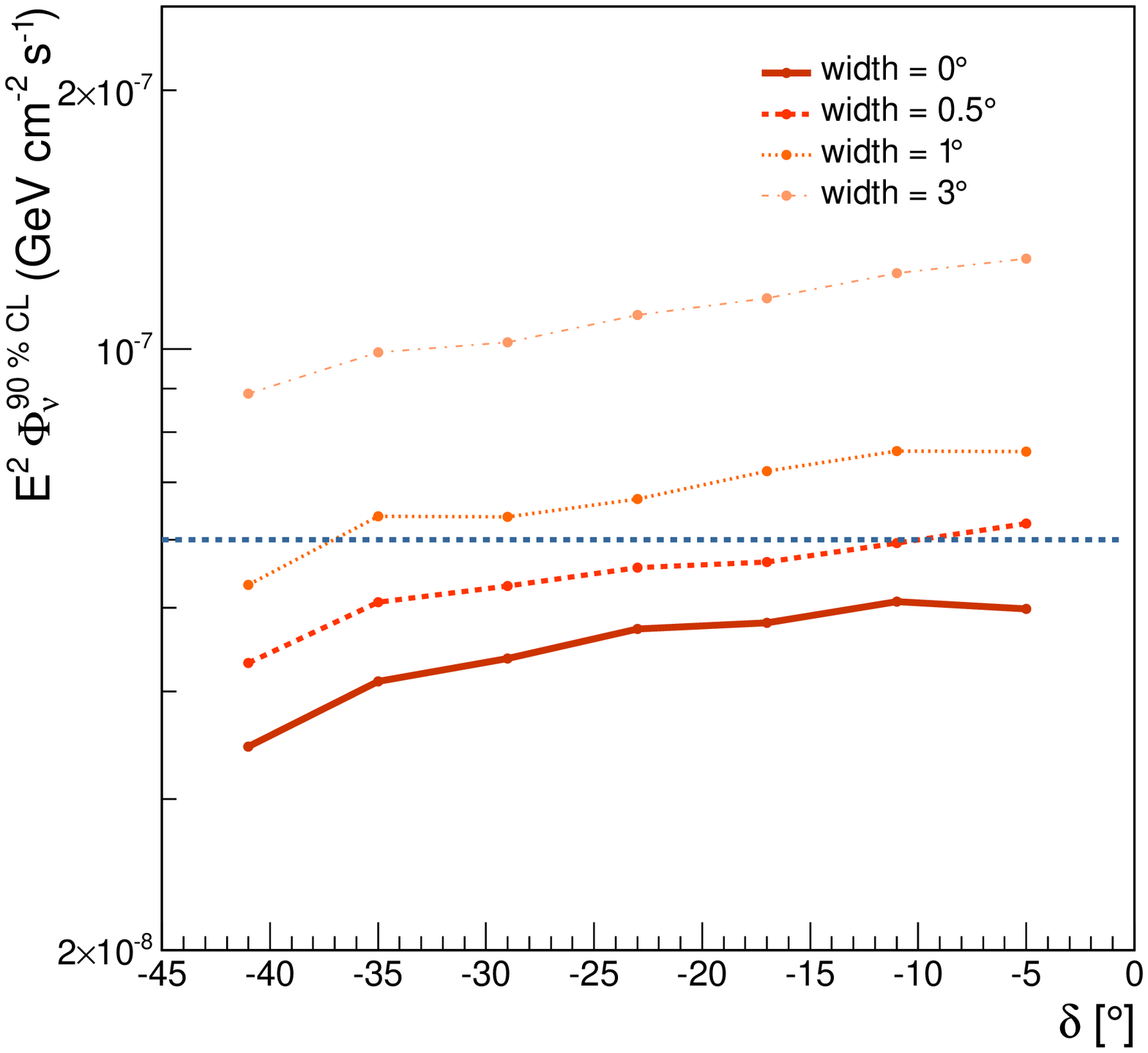}
  \caption{ 90\% C.L. upper limits obtained for different source
    widths as a function of the declination. The blue horizontal
    dashed line corresponds to the signal flux given in ref.~\cite{concha}.}
\label{ICluster}
\end{figure}

The full-sky search looks for an excess anywhere in the part of the
sky visible to ANTARES.  After unblinding, the most significant
cluster found had a 2.7\% post-trial $p$-value (a 2.2$\sigma$
effect). This is not significant enough to claim a signal.  The 90\%
confidence level limits on the muon neutrino flux from point sources
extracted from the absence of a signal are given in declination bands
of 1$^\circ$ by the light blue-dashed line in Fig.~\ref{psources}. A
second search is done using a list of 50 candidate sources (see
ref.~\cite{antpoint}). In this case, the largest post-trial $p$-value
is 6.1\% (1.9$\sigma$) for the candidate source HESS J0632+057. The
limits for these 50 sources are given in Fig.~\ref{psources} by the
large blue squares and the sensitivity is given by the thin blue
line. The corresponding limits and sensitivity are given by the red
squares and thin line, respectively. One is tempted to conclude that
even for part of the southern sky (negative declinations) IceCube has
a better sensitivity, but given their selection method for those
declinations (basically very high energy downgoing events), this limit
only applies for the high energy region, where Galactic sources are
not expected to have a sizeable fraction of their emission. This is
better seen comparing the blue (ANTARES) and red (IceCube)
small-square lines for which the sensitivity is given with the
constraint E$_{\nu} <$100~TeV. While the ANTARES sensitivity
marginally decreases, that of IceCube practically disappears for
southern sky sources.

A point source close to the Galactic Centre has been
proposed~\cite{concha} as a possible explanation to the accumulation
of seven events in its neighbourhood in the first sample of cosmic
neutrinos announced by IceCube~\cite{HESE1}. The corresponding
normalization of the flux of this source would be
6$\times$10$^{-8}$GeV cm$^{-2}$ s$^{-1}$ and would be located around
$\alpha= -$79$^{\circ}$, $\delta= -$23$^{\circ}$.  However, due to the
large error on the direction estimates of these IceCube events, if
present, the location of the source would have a high uncertainty.  We
have performed a search in a region of 20$^{\circ}$ around the
proposed location.  The trial factor of this analysis is smaller than
that of the full-sky search because of the smaller size of the
region. In addition to the point source hypothesis, three
Gaussian-like source extensions are assumed (0.5$^{\circ}$,
1$^{\circ}$ and 3$^{\circ}$).  No significant cluster has been
found. Fig.~\ref{ICluster} shows the 90\% CL flux upper limits
obtained for the spatial extensions of the neutrino source as a
function of the declination. The presence of a point source with a
flux normalization of 6$\times$10$^{-8}$~GeV cm$^{-2}$ s$^{-1}$
anywhere in the region is excluded.  Therefore, the excess found by
IceCube in this region cannot be caused by a single point
source. Furthermore, a source width of 0.5$^{\circ}$ for declinations
lower than $-$11$^{\circ}$ is also excluded. The results are not
affected by a cutoff at energies of the order of PeV, since for an
E$^{-2}$ spectrum the contribution of neutrinos above that energy is
small, of the order of a few percent.

\section{Indirect search for dark matter}
  If dark matter is composed of weakly interacting massive particles
  (WIMPs), these would tend to accumulate by elastic scattering and
  the gravitational pull in the centre of massive astrophysical
  objects like stars, galaxies and clusters of galaxies. There they
  can annihilate and produce standard model particles whose decays
  would produce neutrinos that can be detected at Earth. In some
  cases, e.g. the Sun, the high energy neutrino background expected
  from normal astrophysical processes is very small, so a bunch of
  high energy neutrinos would be enough to claim a signal. ANTARES has
  looked for high energy neutrinos coming from the Sun and the centre
  of our Galaxy.

  The ANTARES search for high energy neutrinos coming from the
  annihilation of WIMPs in the centre of the Sun used data collected
  during 2007 and 2008 corresponding to a livetime of 295 days. The
  search was performed using upgoing neutrino events whose direction
  pointed back to the Sun.  Simulated data was used to optimise the
  cuts on the reconstruction quality of the muon track produced by the
  neutrino and on the angular separation of the neutrino and the Sun
  direction. The simulated data included WIMP annihilation to $b
  \bar{b}$, $W^+ W^-$ and $\tau^+ \tau^-$ without any model
  assumption, i.e. with the same 100\% branching ratios for all the
  channels.  The cuts on the track reconstruction quality and on the
  opening angle of the cone around the Sun for which tracks were
  accepted were obtained optimising the model rejection factor for
  each WIMP mass and each channel. After unblinding, the number of
  selected events was in agreement, within the statistical errors,
  with background expectations. Upper limits at the 90\% confidence
  level on the flux of a neutrino signal as a function of the WIMP
  mass were obtained using Poisson statistics with the Feldman-Cousins
  recipe for all the three channels~\cite{AntDMSun}.  Assuming
  equilibrium between the WIMP capture and self-annihilation rates in
  the Sun, limits on the spin-independent (SI) and spin-dependent (SD)
  WIMP-proton scattering cross-sections can be obtained.

\begin{figure}[ht]
\centering
  \includegraphics[height=0.4\textheight]{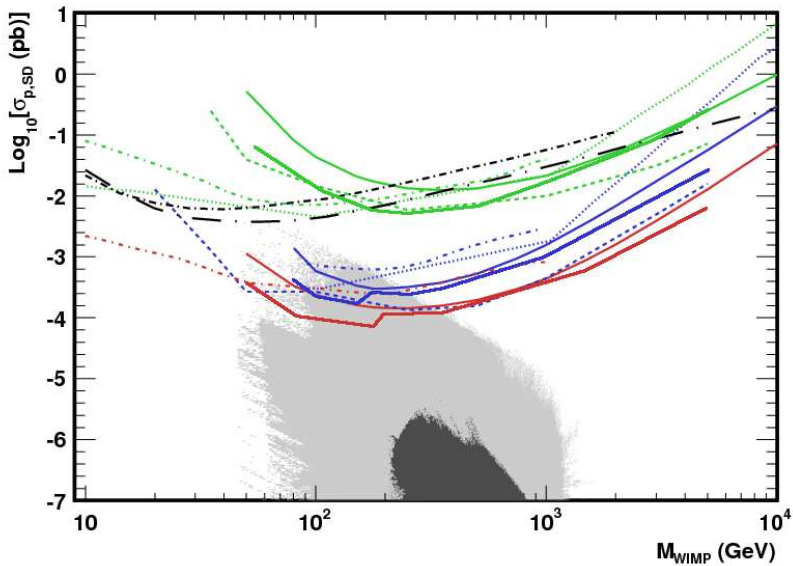}
  \caption{90\% C.L. upper limits on the spin-dependent
    WIMP-proton cross-section as a function of the WIMP mass derived
    from the neutrino flux limits from the Sun for the three
    self-annihilation channels $b \bar{b}$ (green lines), $W^+ W^-$
    (blue lines) and $\tau^+ \tau^-$ (red lines). The type of line
    indicates the experiment which set the corresponding limits: solid
    lines for ANTARES, dashed-dotted for Baksan (1978-2009), dotted
    for Super-Kamiokande (1996-2008) and dashed lines for IceCube-79
    (2010-2011). The black lines correspond to the limits imposed by
    direct search experiments: short dot-dashed lines for SIMPLE
    (2004--2011) and long dot-dashed for COUPP (2010-2011). The dark
    and light grey shaded areas show the results of a grid scan of the
    CMSSM and MSSM-7 SUSY models, respectively. See text for
    references.}
\label{SDDM}
\end{figure}

  Fig.~\ref{SDDM} shows the limits for the spin-dependent
  WIMP-proton cross-section, case in which the limits imposed by
  neutrino telescopes are in general very competitive, because the
  capture rate is very sensitive to this cross-section since the Sun
  is mostly composed of protons.  The colour of the curves in the
  figure indicate the channel, green for $b \bar{b}$, blue for $W^+
  W^-$ and red for $\tau^+ \tau^-$. The type of line indicates the
  experiment which set the corresponding limits: solid lines for
  ANTARES (2007-2008), dashed-dotted for Baksan
  (1978-2009)~\cite{baksan}, dotted for Super-Kamiokande
  (1996-2008)~\cite{SK} and dashed lines for IceCube-79
  (2010-2011)~\cite{DMIC}. The black lines correspond to the limits
  imposed by direct search experiments: short dot-dashed lines for
  SIMPLE (2004--2011)~\cite{SIMPLE} and long dot-dashed for COUPP
  (2010-2011)~\cite{COUPP}. Also shown are the results of a grid scan
  of the CMSSM and MSSM-7, the dark and light grey shaded areas,
  respectively. As can be seen, ANTARES limits are very competitive
  with respect to other experiments and skim the region predicted by
  MSSM-7.

  As in the case of the Sun, WIMPs can also accumulate in the centre
  of the Galaxy and annihilate producing neutrinos. We have used the
  data recorded by ANTARES between 2007 and 2012, corresponding to a
  livetime of 1321 days, to search for neutrinos coming from the
  Galactic Centre.  The method is similar to that of the Sun, but in
  this case two reconstruction strategies that complement each other
  in different WIMP mass ranges have been used.  Events reconstructed
  in one single line have also been used: these lack the measurement
  of the azimuth angle and therefore the expected background increases
  slightly, but this is compensated by the increase in statistics.  As
  in the previous analysis the cuts on the quality parameter --given
  in this case by one or the other of the reconstruction strategies--
  and on the angular separation of the track with respect to the
  position of the Galactic Centre are optimised using a model
  rejection factor. After unblinding, the number of events found are
  in agreement with those expected for the background and thus limits
  are imposed on the neutrino fluxes for each of the different
  annihilation channels.  A useful quantity to compare among
  experiments that use different techniques to detect dark matter is
  the velocity averaged WIMP self-annihilation cross-section,
  $<\sigma_A \cdot \textrm{v}>$. To extract this quantity from the
  limits on the neutrino flux, some assumptions have to be
  made. First, a distribution has to be assumed for the density
  profile of the dark matter in the Galaxy, $\rho_{DM}$, that enters
  into the flux estimation through the so-called J-factor, i.e.  the
  integral along the line of sight of the square of the WIMP density:

\begin{equation}
J(\Delta \Omega) = \int_{\Delta \Omega} \int \rho^2_{DM} \, dl \, d\Omega
\end{equation}

\begin{figure}[ht]
\centering
  \includegraphics[height=0.4\textheight]{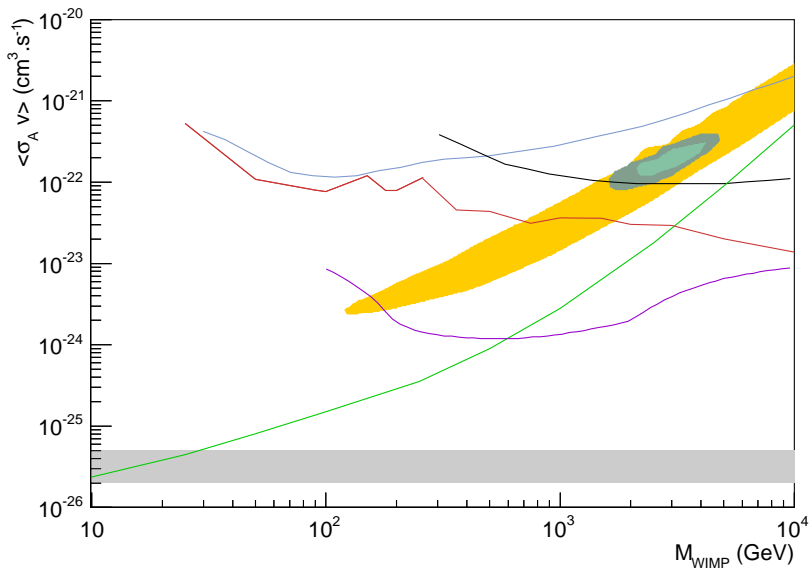}
  \caption{90\% C.L. upper limit set by ANTARES on
    $<\sigma_A \cdot \textrm{v}>$ obtained for the reference $\tau^+
    \tau^-$ channel (red line). The following limits are also shown:
    IceCube--DeepCore 79 (2010-2011) for the Galactic Centre (blue
    line), IceCube--59 (2009-2010) for the Virgo cluster (black line),
    Fermi-LAT (2008-2010) for the analysis of 10 satellite galaxies
    (green line) and MAGIC (2011-2013) for Segue~1 (purple line).  The
    regions favored by PAMELA (orange area) and by PAMELA, Fermi-LAT
    and H.E.S.S. (green ellipses) interpreted as dark matter
    self-annihilations are also shown.  The gray band indicates the
    natural scale for which all the dark matter is considered to be
    composed of WIMPs only. See text for references.  }
\label{sigmav}
\end{figure}

We use the Navarro-Frenk-White galactic dark matter halo
profile~\cite{NFW}, with a parameter r$_s$=21.7 kpc and normalised in
such a way that at the Sun's position the density is
0.4~GeV cm$^{-3}$. In Fig.\ref{sigmav}, the red line
indicates our 90\% confidence level upper limit on $<\sigma_A \cdot
\textrm{v}>$ obtained for the reference $\tau^+ \tau^-$ channel.  For
the sake of comparison, the following limits on $<\sigma_A \cdot
\textrm{v}>$ are also given: IceCube--DeepCore 79 (2010-2011) for the
Galactic Centre indicated by the blue line~\cite{ICGC}, IceCube--59
(2009-2010) for the Virgo cluster given by the black
line~\cite{ICVirgo}, Fermi-LAT (2008-2010) for the joint analysis of 10
satellite galaxies given by the green line~\cite{Fermi} and MAGIC
(2011-2013) for Segue~1 given by the purple line~\cite{Magic}.  The
regions favored by PAMELA (orange area) and by PAMELA, Fermi-LAT and
H.E.S.S. (green ellipses) interpreted as dark matter
self-annihilations are also shown~\cite{various}.  The gray band
indicates the natural scale for which all the dark matter is
considered to be composed of WIMPs only.

\section{Summary}
The ANTARES neutrino telescope started to take data in 2007.  The
search for a neutrino diffuse flux has not yielded a neutrino signal
and an upper limit on such a flux has been set. Let us note that the
sensitivity of ANTARES for neutrino diffuse fluxes is not very far
from the signal observed by IceCube, although still insufficient to
confirm it. ANTARES is the most sensitive neutrino telescope for point
sources in the southern sky, especially for energies lower than
100~TeV, where most of the neutrino events from a Galactic source are
expected to lie.  ANTARES can exclude a point source as the origin of
the cluster of events observed by IceCube not far from the Galactic
Centre. Even a $0.5^{\circ}$ extended source is excluded for most of
the declinations near the Galactic Centre. Concerning the indirect
search for dark matter using neutrinos, ANTARES has been able to set
very competitive limits on the flux of neutrinos coming from WIMP
self-annihilation in the Sun and the Galactic Centre.

\section{Acknowledgments}
   The author wants to thank the organizers for inviting him to give a
   talk at the Neutrino 2014 Conference. The exciting scientific atmosphere
   of the conference and the charming city of Boston were a delightful 
   experience.  
   He also gratefully acknowledges the support of the
   Spanish MINECO through project FPA2012-37528-C02-01, the MultiDark
   Consolider Project, CSD2009-00064, and the Generalitat Valenciana
   via Prometeo-II/2014/079.


\end{document}